\documentclass[aps,twocolumn]{revtex4}
\usepackage[dvips]{color}
\usepackage{epsfig}
\usepackage[novolumeabbr,liter]{unitsdef} 

\begin{document}

\title{$sp^{2}$/$sp^{3}$ carbon ratio in graphite oxide with different preparation times}

\author{D.~W.~Lee\renewcommand{\thefootnote}{${*}$}\footnote{Corresponding author. E-mail: dongwookleedl324@gmail.com. Phone: +65 6513 8459; Fax: +65 6795 7981.}$^{,}$\renewcommand{\thefootnote}{${\dag}$}\footnote{University of Cambridge}$^{,}$\renewcommand{\thefootnote}{${\ddag}$}\footnote{Nanyang Technological University} and J.~W.~Seo$^{\dag,\ddag}$
}

\affiliation{Cavendish Laboratory, University of Cambridge, JJ Thomson Ave.,Cambridge CB3 0HE, United Kingdom}
\affiliation{Division of Physics and Applied Physics, Nanyang Technological University, 637371 Singapore}

\date{\today}
\begin{abstract}

Graphite oxide is an amorphous insulator. Although several models have been suggested, its structure remains controversial. To elucidate this issue, 5 samples were prepared by the Brodie process and the Staudenmaier process. The electronic structure of graphite oxide was examined with x-ray absorption near edge structure and the ratio of $sp^2$ to $sp^3$ bonded carbon atoms was investigated with x-ray photoemission spectroscopy as a function of sample preparation times. It was found that this ratio approaches 0.3 exponentially with a characteristic time of 1.5 weeks. We believe this long characteristic time is the reason the structure has remained unclear.

\end{abstract}

\maketitle

\section*{1. Introduction}

Graphite, which is made up of graphene layers, has been utilized by scientists in recent decades for adsorption due to its large surface area or as a host for intercalation. Although CO and CO$_{2}$ are produced during its oxidation in air, it becomes amorphous insulating graphite oxide (GO) if the oxidation is carefully controlled. A layered structure remains after oxidation with structural distortions due to the coexistence of $sp^{2}$- and $sp^{3}$-bonded carbon atoms \cite{Scholz,Lerf}. Recently, GO has attracted attention because it can be used to manufacture graphene at low cost \cite{DAN} and reduced GO can be used as a transparent and flexible electronic material \cite{GOKI}. In addition, the dielectric constant of NaOH reacted GO shows frequency- and temperature dependent behavior \cite{DOW1}. However, despite these uses, its exact structure is unknown.

\begin{figure}[!b]
\epsfig{file=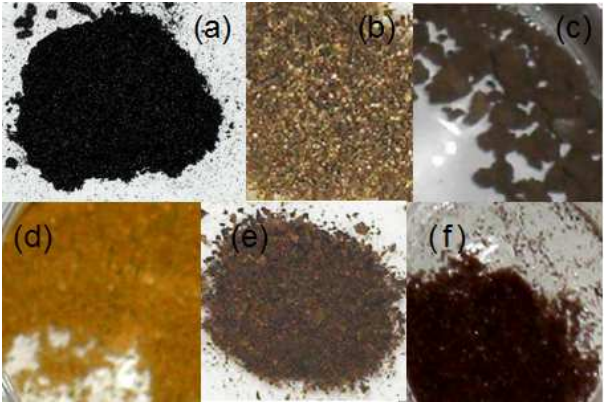, width=8.0cm}
\caption{(Color online) \textbf{Sample pictures of GO.} (a) graphite (b) GO1 (c) GO2 (d) GO3 (e) GO4 (f) GO5.}
\end{figure}

Since it was first discovered in 1859 by Brodie \cite{Brodie}, four different preparation methods have been developed: the Brodie process \cite{Brodie}, the Staudenmaier process \cite{Staudenmaier}, the Hummers process \cite{Hummer0,Hummer} and anodic oxidation of graphite electrodes in nitric acid \cite{Hudson-Hunter}. Five main different structural models have been proposed for GO: Hofman's model \cite{Hofmann}, Ruess's model \cite{Ruess}, Sholtz \& Boem's model \cite{Scholz}, Szab\'{o}'s model \cite{Szabo}, and Lerf \& Klinowski's model \cite{Lerf}. Hofmann first proposed that only epoxy (-O-) groups are situated on the surface. Although Hofmann's model explained the existence of epoxy (-O-) groups, it does not include hydrogen containing groups. Ruess assumed that the carbon sheets were wrinkled and they consisted of trans-linked cyclohexane chairs. His was the first model to account for the hydrogen content in GO. Ruess' model was revised by Scholz and Boem who assumed that there were no ether groups in GO. Scholtz \& Boem's model was adapted to incorporate ketone groups by Szab\'{o}. Lerf and Klinowski proposed the existence of hydroxyl (-OH) groups in addition to epoxy groups. In addition to these five models, some models based on experiments and simulation \cite{Boukhvalov,DOW2} have been proposed over the last few years.

\begin{table}[!b]
\begin{tabular}{|c|c|c|c|c|c|} \hline
sample & GO1 & GO2 & GO3 & GO4 & GO5 \\ \hline
preparation time (week) & 3 & 4 & 6 & 5 & 10   \\ \hline
preparation method & B & B & B & S & S\\ \hline
\end{tabular}\caption{Sample preparation time for GO samples. B stands for Brodie and S stands for Staudenmaier.}
\end{table}

Recent researches on GO have been performed by X-ray photoemission spectroscopy (XPS) \cite{Galuska,DOW2,DOW3,Hontora-Lucas}, X-ray absorption near edge structure (XANES) \cite{DOW3,YHLEE}, solid-state Nuclear magnetic resonance (NMR)\cite{Lerf,Szabo,DOW2}, and etc. Most models including the above agree that GO is amorphous and insulating. However, they suggest different structures because they did not consider different preparation time. Thus, it is worthwhile to try to understand why doubts about the structure still remain. This work therefore considers the question: does GO really have so many different structures or does its structure depend on the preparation time?

\section*{2. Experimental}

For these experiments, the GO samples were prepared by the Brodie process \cite{Brodie} and the Staudenmaier process \cite{Staudenmaier}. The Brodie process is as follows. 5.0\gram of graphite (99.995+\percent purity, 45\micm, Aldrich) was added into 62.5\milliliter of fuming nitric acid. After cooling this mixture in an ice bath, 25.0\gram of potassium chlorate was slowly added. After the mixture had reached room temperature, it was placed in a water bath, heated slowly to a temperature of 45\celsius and kept at this temperature for 20\hour. Subsequently, the mixture was poured into 125\milliliter of cold distilled water, warmed to 70\celsius, and then centrifuged, decanted, and dried overnight at 70\celsius. The oxidation process was performed 2 times per week. The whole process were repeated. It took from 3 weeks to 6 weeks to prepare GO samples by the Brodie method. The Staudenmaier process is as follows. 5.0\gram of graphite (99.995+\percent purity, 45\micm, Aldrich) was added to fuming nitric acid (25\milliliter) and sulphuric acid (50\milliliter). After cooling the mixture down to 5\celsius in an ice bath, 25.0\gram of potassium chlorate was slowly added to the solution while stirring. The mixture was kept at this temperature for 3 days. It was then transferred into 1$\ell$ of water. The solution was immediately filtrated and dried several times. The oxidation process was performed 2 times per week. The whole process were repeated. The GO samples took 5 and 10 weeks to be prepared by the Staudenmaier method (2 times of the oxidation per week). Five samples were prepared: GO1, GO2, and GO3 by the Brodie method, and GO4 \& GO5 by the Staudenmaier method.

\begin{figure}[!t]
\epsfig{file=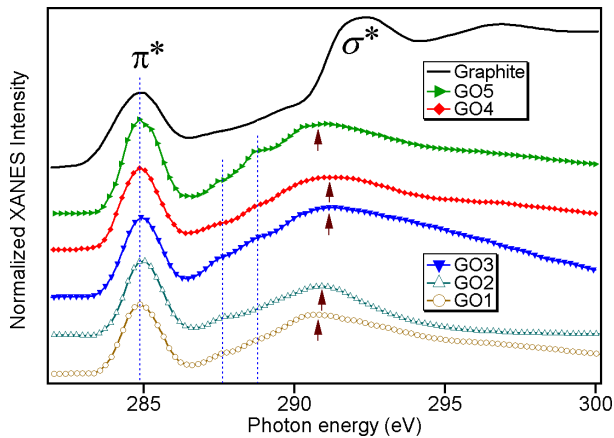, width=8.0cm}
\caption{(Color online) \textbf{Characterization of the samples using C \emph{K}-edge XANES spectra.}}
\end{figure}

\begin{figure}[!t]
\epsfig{file=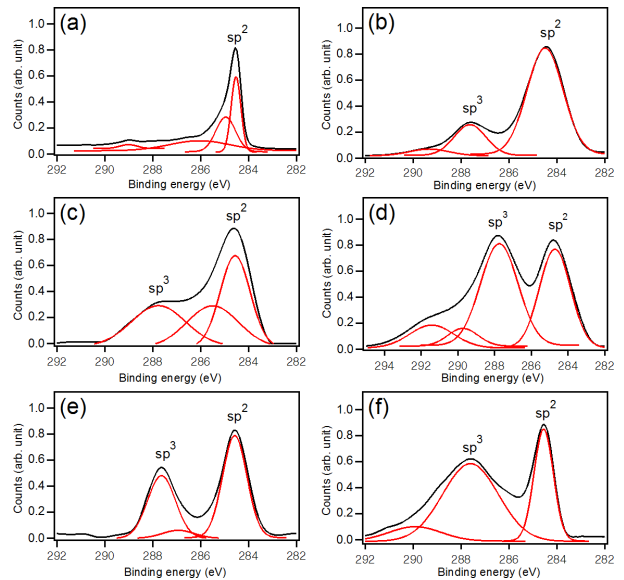, width=8.0cm}
\caption{\textbf{Characterization of the samples using C 1s XPS spectra.} (a) graphite (b) GO1 (c) GO2 (d) GO3 (e) GO4 (f) GO5.}
\end{figure}

GO samples were characterized by XPS and XANES, which were performed on the BACH beamline at Elettra in Italy. Its radiation source is based on two APPLE-II helical undulators. The storage ring was operated with an electron energy of 2.0\GeV and a current was between 160 and 270\mA. GO powder samples were mounted on an oxygen free high-conductive copper (OFHC) plate with a silver epoxy (Dupont 4929) embedded. After mounting GO samples, the chamber was baked out for 9 hours at 100\celsius. The chamber was cooled down again. The XPS and XANES experiments were conducted when the pressure inside the chamber reached high 10$^{-10}$ torr. XPS spectra were obtained using a 150 mm VSW hemispherical electron analyzer with a 16-channel detector at room temperature and the incident photon energies were calibrated by measuring the Au 4f photoelectron core level. 388\eV photon was used for XPS. The pass energy of the analyzer was 40\meV and the resolution was 100\meV. The XANES spectra were acquired in the total electron yield (TEY) mode at room temperature.

\begin{table}[!b]
\begin{tabular}{|c|c|c|c|c|c|} \hline
 sample  & GO1 & GO2 & GO3 & GO4 & GO5 \\ \hline
$sp^{2}$ : $sp^{3}$ & 3.95 & 2.22 & 0.70 & 1.25 & 0.50 \\ \hline
\end{tabular}\caption{The ratio of $sp^{2}$- to $sp^{3}$-bonded carbon atoms in the GO samples. This ratio is determined from the areas of Gaussian fits to the peaks in the XPS data in Figure 3.}
\end{table}

\section*{3. Results and discussion}

Figure 1 shows the samples. Graphite (Figure 1(a)) is black. Although the GO samples look different from graphite, they also look slightly different from each other. GO1 is brown (Figure 1(b)), GO3 is yellow (Figure 1(d)) while GO2, GO4 and GO5 are dark brown. Table 1 shows the sample preparation times, which vary from 3 to 10 weeks. GO3 looks more yellow than the other samples prepared by the Brodie method. GO5 is slightly brighter than GO4 even though they were both prepared by the Staudenmaier method. Figure 1 and Table 1 reveal that the samples become brighter as the preparation time increases.

\begin{figure}[!t]
\epsfig{file=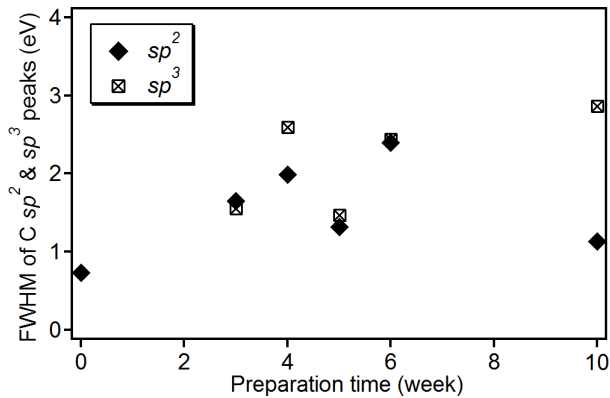, width=8.0cm}
\caption{\textbf{FWHM of C $sp^{2}$ and $sp^{3}$ peaks in Figure 3.}}
\end{figure}

Figure 2 shows the C \textit{K}-edge XANES spectra of the polycrystalline graphite. The spectra of the samples were normalized by the intensity of the peak at 285\eV which is the $\pi^{*}$ state of C=C ($1s \longrightarrow \pi^{*}$ excitations) \cite{Joachim,Pianetta}. Graphite has the peak at 293\eV which is assigned to $\sigma^{*}$ bonds of C=C ($1s \longrightarrow \sigma^{*}$ transitions)  \cite{Joachim,Pianetta}. The peak from $\sigma^{*}$ bonds of C=C in GO samples is shifted to 291\eV (the arrows in Figure 2). GO samples have two more peaks at 288\eV and 289\eV. The peak at 288\eV refers to $\pi^{*}$ bonds of C-OH \cite{Francis}. The peak from hydroxyl groups at 288\eV are formed from the initial stage (GO1) and its peak continues to grow. The peak at 289\eV is from $\pi^{*}$ bonds of C-O-C which are vertically aligned \cite{DOW3,Hennies}, that is, epoxy groups. The peak intensity from the $\sigma^{*}$ state in GO samples increases as the preparation time increases. The longer reaction times takes, the more C=C bondings in GO become decomposed, suggesting disorder in the $\pi$ states, that is, localized C=C bondings are formed \cite{Fink,Robertson}. As a result, $\sigma^{*}$ peaks becomes stronger than that of $\pi$-bonding in C=C. And the increase of localized C=C will lead to the change in the ratio of $sp^{2}$ to $sp^{3}$ carbon. Although we do not measure the mechanical, thermal, and electrical properties of GO, these properties will be influenced by the change in  the ratio of $sp^{2}$ to $sp^{3}$ carbon. If there is only $sp^{2}$ carbon in graphite, there is no bending and warping. However, if $sp^{2}$ and $sp^{3}$ carbon coexist, bending and warping appear. As appear more $sp^{3}$ carbons in GO, more bending and warping appears. Longer preparation time makes GO more defective.

XPS spectra of the six different samples are exhibited in Figure 3. Graphite has two peaks in Figure 3(a). The peak at 284.5\eV is assigned to C-C bonds \cite{Javier} and is asymmetric due to defragments such as anthracene \cite{Estrade}. The peak at 289\eV is from plasmon \cite{Xie} which is a collective behavior of delocalized valence electrons of graphite. GO samples show two main peaks in Figure 3(b) - Figure 3(f). The peak at 284.5\eV is assigned to carbon atoms with $sp^{2}$ hybridized orbitals \cite{Javier}. The other peak originates from C-O in alcohol with $sp^{3}$ hybridized orbitals \cite{Cheung,Katzman,Hiroki}. GO3 has three main peaks unlike other GO samples. The second peak at 285.5\eV is due to structural defect with $sp^{2}$ hybridized orbitals \cite{Estrade}. The peaks in Figure 3 were fitted with Gaussian functions to determine their areas. Dividing these areas gives the ratio of the number of $sp^{2}$ to $sp^{3}$ bonds (Table 2). Comparison with Table 1 shows that this ratio is a monotonic function of preparation time, regardless of the method used. That is, if the sample is prepared for longer, more reactions occur and so more chemical groups form in the sample. In addition, it is deduced that the chemical groups with $sp^{2}$ hybridized orbitals such as epoxy groups are formed at initial stages and then the chemical groups with $sp^{3}$ hybridized orbitals such as the hydroxyl groups appear to form.

\begin{figure}[!t]
\epsfig{file=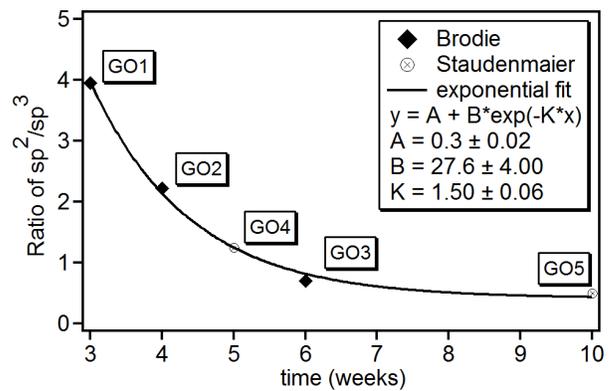, width=8.0cm}
\caption{\textbf{The ratio of $sp^{2}$/$sp^{3}$ orbitals in samples vs sample preparation time.} The ratio of carbon atoms with $sp^{2}$ orbitals to carbon atoms with $sp^{3}$-hybridized orbitals in GO samples as a function of sample preparation time.  The solid line is a least square fit of an exponential function.}
\end{figure}

Figure 4 shows the full width at half maximum (FWHM) of the peaks in Figure 3. The FWHM of $sp^{2}$ peaks ranges from 0.73\eV to 2.4\eV. The FWHM of $sp^{3}$ peaks ranges from 1.56\eV to 2.87\eV. A possible origin for the peak broadening is likely to be the charging. However, this can be ruled out because the charging of the surface should give rise to homogeneous peak broadening for all the peaks. In our data, we observe different values for the peak broadening. Another explanation for the broadening is due to the structural disorder that accompanies the oxidation of graphite\cite{Galuska}. As the preparation time increases, more $sp^{2}$ bonds are decomposed into $sp^{3}$ bonds due to the reaction. Therefore, the FWHM of $sp^{2}$ increases with the preparation times. However, decomposition of $sp^{2}$ bonds are likely to become slow after 6 weeks and the FWHM decreases to 1.13 (GO5), while more $sp^{3}$ bonds are generated with the preparation time and GO samples grow more defective.

Figure 5 displays the ratio of $sp^{2}$ carbon atoms to $sp^{3}$ carbon atoms in the GO samples (Table 2) against their preparation time (Table 1). The line in Figure 5 shows the least square fit of an exponential function.  This shows that the ratio asymptotically approaches 0.3 $\pm$ 0.02, with a characteristic decay time (K in Figure 5) of 1.5 weeks.  This suggests that it is not possible to prepare GO samples with only $sp^{3}$-hybridized orbitals irrespective of the preparation time. This is likely to be because the localized double bonds in GO are so stable that all of them are not destroyed by the reaction with strong acids such as nitric acid and sulfuric acid.

Since GO is amorphous, it tends to continue to react with chemicals during preparation. Unlike crystalline materials, atoms in amorphous materials move and interact over time and have superior catalytic activity (amorphous materials are used as catalysts for this reason \cite{Schwarz,Li-H,W-wang}). We believe that the long characteristic time scale of the preparation time (almost one and a half weeks) is the reason for the ambiguity of its structure.

\section*{4. Conclusions}

We synthesized various GO samples with different preparation times and investigated them with XANES and XPS. The XANES data reveal that GO samples have chemical groups on the surface such as hydroxyl groups (-OH) and epoxy groups (C-O-C). The peak from $\sigma^{*}$ bond grows with the preparation time. The XPS spectra show that GO samples have two kinds of carbon atoms: carbon atoms with $sp^{2}$-hybridized orbitals and carbon atoms with $sp^{3}$-hybridized orbitals. The ratio was plotted against the sample preparation time. The ratio decreases as it takes longer time to prepare the samples. The ratio approaches 0.3 exponentially with a characteristic decay time of 1.5 weeks.

\section*{Acknowledgements}

The authors are grateful to L.~M.~Brown and G.~R.~Jelbert for helpful discussions. The authors are also indebted to BACH beamline staffs.

\clearpage

\end{document}